\documentclass[10pt]{article}
\usepackage{a4wide,bm,cite}
\usepackage{epsfig}
\usepackage{amsmath}
\usepackage{amssymb}
\DeclareMathOperator{\sech}{sech}
\begin{document}
%\begin{frontmatter}
\title{{\bf{Electromagnetic breathing dromion-like structures in an anisotropic ferromagnetic medium}}}% Force line breaks with \\

\author{\bf \ Sathishkumar Perumal$^{a,}$\thanks{Corresponding author and E-mail address: perumal$\_$sathish@yahoo.co.in, Tel:+91-4288-274213, Fax:+91-91-4288-274757}, \bf J. Sivapragasam$^{a, b}$ and M. Lakshmanan$^{c}$ \\
{\footnotesize $^{a}$ Department of Physics, Sri Paramakalyani College, Alwarkurichi - 627 412, Tenkasi District, Tamil Nadu, India.}\\
{\footnotesize $^{b}$ Department of Physics, K.S.R. College of Engineering, Tiruchengode-637 215, Tamil Nadu, India.}\\
{\footnotesize $^{c}$ Department of Nonlinear Dynamics, School of Physics, Bharathidasan University, Tiruchirapalli-620 024, Tamil Nadu, India}}
\date{}
\maketitle
\rule{5.8in}{0.2pt}
\begin{abstract}
$~~~~~$ The influence of Gilbert damping on the propagation of electromagnetic waves (EMWs) in an anisotropic ferromagnetic medium is investigated theoretically. The interaction of the magnetic field component of the electromagnetic wave with the magnetization of a ferromagnetic medium has been studied by solving the associated Maxwell's equations coupled with a Landau-Lifshitz-Gilbert (LLG) equation. When small perturbations are made on the magnetization of the ferromagnetic medium and magnetic field along the direction of propagation of electromagnetic wave by using the reductive perturbation method, the associated nonlinear dynamics is governed by a time-dependent damped derivative nonlinear Schr\"odinger (TDDNLS) equation. The Lagrangian density function is constructed by using the variational method to solve the TDDNLS equation to understand the dynamics of the system under consideration. The propagation of EMW in a ferromagnetic medium with inherent Gilbert damping admits very interesting nonlinear dynamical structures. These structures include Gilbert damping-managing symmetrically breathing solitons, localized erupting electromagnetic breathing dromion-like modes of excitations, breathing dromion-like soliton, decaying dromion-like modes and an unexpected creation-annihilation mode of excitations in the form of growing-decaying dromion-like modes.\\[5mm] 
\rule{5.8in}{0.2pt}
\end{abstract}
%\end{frontmatter}
\section{Introduction}

The magnetization dynamics associated with the propagation of intense electromagnetic (EM) waves in nonlinear ferromagnetic media have attracted much attention from the perspectives of fundamental physics as well as technological applications like magnetic recording, fast data retrieval, and high density data storage devices [1-3]. When an EM wave propagates through a nonlinear ferro, ferri or anti-ferromagnetic medium, the interplay between the magnetic field components of EM wave and the magnetization of magnetic medium results in several interesting phenomena such as harmonic generation, self-focusing, domain wall propagation, EM soliton propagation and breather-like solitary excitations [4-8]. The nonlinear modes of excitations are described by the families of nonlinear dynamical equations that include nonlinear Schr\"odinger (NLS) equations for nonlinear spin excitations in the form of solitons, derivative nonlinear Schr\"odinger (DNLS) equation for breather-like solitary excitations, modified Korteweg-de-Vries (mKdV) equations for domain wall-type excitations and Korteweg-de-Vries (KdV) equations for polarisation in the long-wave approximation [9-14]. Further, it has been observed that the plane EM wave and the magnetization of the medium are modulated in the form of an EM soliton when it propagates through a charge-free, isotropic, and anisotropic ferromagnetic medium [15-17]. When free charges are present in the medium, however, the EM soliton slows down and the amplitude diminishes and becomes damped [18]. Moreover, the propagation of EM waves in an antiferromagnetic medium with Dzyaloshinskii-Moriya interaction has been studied theoretically and it was found that the propagation of EM waves is governed by breather-like solitary excitations [19]. 

The study of the breathing nature of the soliton excitations in the various physical systems has received a lot of attention over the past few years. Breathers are recognized as spatially localized nonlinear excitations in the nonlinear systems, i.e., solitary waves can breathe periodically with virtually undamped long-lived oscillations of amplitude and width. The interesting features of breathing soliton are that it does not damp down and no radiation is emitted [12, 20]. The breathing mode of soliton excitations has been experimentally observed in a quasi one-dimensional antiferromagnetic spin chain with magnetic field [21]. It has been shown that a classical anisotropic Heisenberg discrete spin chain with and without an external magnetic field can have a variety of classes of exact soliton solutions expressed in terms of Jacobian elliptic functions [22]. The stability and explicit analytical solutions of one, two and three spins excitations for the anisotropic Heisenberg spin chain with onsite anisotropy and constant external magnetic field have been studied in [23]. However, further work is needed in this direction to identify such interesting class of soliton solutions in a classical anisotropic Heisenberg spin chain.

Soliton explosions are the most striking and fascinating nonlinear dissipative phenomena studied in fiber laser, Frenkel-Kontorova model with anharmonic interatomic interactions and physical systems with nonlinear damping [24, 25]. Soliton explosions are unstable solutions caused by instabilities and their mechanisms have been studied in generalized Klein-Gordon equations with nonlinear damping and spatio-temporal perturbations [26]. The space-dependent perturbations and nonlinear damping that periodically create instability in the location where the soliton is located at that instant result in a highly non-stationary spatio-temporal state in which the soliton is not allowed to recover its original shape, leading to soliton explosion. In practice, it is realized that such mechanisms can account for some of the events that have lately been reported to happen in heart tissue and a transmission line at a Josephson junction [27, 28].

In this study, we demonstrate the existence of localized electromagnetic breathing dromion-like structures, dromion explosions and decaying breathing dromion-like modes of excitation in an anisotropic ferromagnetic medium with Gilbert damping when subjected to a varying EM wave. In order to put forward the TDDNLS equation for understanding the evolution of the magnetization of the medium, we apply the multiple-scale perturbation method to the system in which almost all explicit parameters appearing in the equations are redefined. The present work is organized as follows. Section 2 introduces the governing dynamical equations. Section 3 employs a reductive perturbation technique that yields the TDDNLS equation. The magnetization components of an anisotropic ferromagnetic medium are constructed by using the variational method with the Lagrangian density function in Section 4. Section 5 provides a summary of the results.

\section{Model and dynamics}

~~~~~The magnetization dynamics of a classical continuum anisotropic ferromagnetic medium with Gilbert damping in the presence of EM field is governed by the Landau-Lifshitz-Gilbert equation [29], 
\begin{equation}
\frac{\partial{\bf{M}}}{\partial t}=\gamma\bigl[{\bf{M}}\times{\bf{F}_{eff}}\bigr]+\Lambda\bigl[{\bf{F}_{eff}}-({\bf{M}}.{\bf{F}_{eff}}){\bf{M}}\bigr], ~~~~~{\bf{M}}^2=1,
\end{equation}
where $ \bf{M(r, t)}$ represents the magnetization ${\bf{M}}=(M^x, M^y, M^z)$ of the ferromagnetic medium. The effective field ${\bf{F}_{eff}}$ in the presence of a varying magnetic field $\bf{H}$ as the magnetic field component of the EM wave propagating through the anisotropic ferromagnetic medium along the z-direction is written as ${\bf{F}_{eff}}=J\nabla^{2}{\bf{M}}-2AM^{z}{\bf{\hat{n}}}+2\beta \bf{H}$, where $J$ is exchange integral which arises due to spin-spin exchange interaction and ${\bf{\hat{n}}}=(0,0,1)$. The anisotropy parameter $A$ describes the strength of the crystal field anisotropy along the z-direction as the easy axis of magnetization, and the contribution due to interaction of the magnetization with the external magnetic field is represented by the term proportional to $\beta=\gamma\mu_B$ in which $\gamma$ and $\mu_B$ represent the gyromagnetic ratio and the Bohr magneton, respectively. The parameter $\Lambda$ is designated as $\gamma\lambda$, where $\lambda$ is the dimensionless Gilbert damping parameter which stands for nonlinear spin relaxation phenomenon. Eq. (1) is equivalent to the Landau-Lifshitz-Gilbert (LLG) equation in which the first term depicts the precession of the magnetization vector about the effective field ${\bf{F}_{eff}}$ with angular frequency ${\bf{\omega}}=-\gamma |{\bf{F}_{eff}}|$. The second term in Eq. (1) represents the Gilbert damping torque which describes the relaxation of magnetization towards the direction of the effective field ${\bf{F}_{eff}}$ for a positive value of $\lambda$ i.e., it controls the rate at which the magnetization vector relaxes to equilibrium state. For negative value of $\lambda$, the magnetization ${\bf{M}}$ spirals away from the direction of ${\bf{F}_{eff}}$ into a direction opposite to ${\bf{F}_{eff}}$ thereby gaining energy and corresponds to switching of magnetization. The Gilbert damping plays a key role in the spin dynamics of magnetic systems. It is commonly assumed that the origin of the Gilbert damping is the spin-orbit coupling and two-magnon scattering process [30, 31]. Materials with large spin-orbit coupling exhibit strong interactions between their magnetic moments and the surrounding environment, resulting in high damping of the magnetization dynamics. Therefore, the processes of magnetization relaxation and switching in spintronics devices are profoundly affected by Gilbert damping. In particular, it affects the velocity of domain-wall in current-carrying domain-wall structures where fast propagation of domain-walls is essential for applications in high-speed spintronic devices such as magnetic race track memory [32]. Now, the evolution of magnetization density associated with an anisotropic ferromagnetic medium where the EM wave propagates through it along the z-direction can be redefined by making use of expression for the effective field ${\bf{F}_{eff}}$. The LLG equation demands that the length of the magnetization vector does not change with time ($\bf{M}^{2}=1$). The one-dimensional form of the evolution of magnetization density after substituting the effective field ${\bf{F}_{eff}}$ in the resultant equation is expressed as   
\begin{eqnarray}
\frac{\partial{\bf{M}}}{\partial t}={\bf{M}}\times \Bigl[J\frac{\partial^2{\bf{M}}}{\partial z^2}-2AM^{z}{\bf{\hat{n}}}+2\beta {\bf{H}}+\lambda\bigl[J\frac{\partial^2{\bf{M}}}{\partial z^2}-2AM^{z}{\bf{\hat{n}}}+2\beta {\bf{H}}\nonumber\\-J({\bf{M}}.\frac{\partial^2{\bf{M}}}{\partial z^2}){\bf{M}}+2A({\bf{M}}.M^{z}{\bf{\hat{n}}}){\bf{M}}-2\beta ({\bf{M}}.{\bf{H}}){\bf{M}}\bigr]\Bigr],
\end{eqnarray} 
where $z$ is now the propagation direction. The magnetization $\bf{M}$ of the ferromagnetic medium is related to the magnetic induction $\bf{B}$ and the magnetic field $\bf{H}$ as ${\bf{B}}=\mu(\bf{H}+\bf{M})$, where $\mu$ is the magnetic permeability of the medium. The variations of the electric field $\bf{E}$ and magnetic field $\bf{H}$ components of an EM wave as it propagates through a ferromagnetic medium in the absence of static and moving charges are governed by the Maxwell equations as 
\begin{equation}
\nabla \cdot \bf{E}=0,
\end{equation}
\begin{equation}
\nabla \cdot \bf{B}=0,
\end{equation}
\begin{equation}
\nabla\times \bf{E}=-\frac{\partial \bf{B}}{\partial t},
\end{equation}
\begin{equation}
\nabla\times \bf{H}=\frac{\partial \bf{D}}{\partial t}.
\end{equation}
The vector representations of the electric field, magnetic field and magnetic induction are ${\bf{E}}=(E^x,E^y,E^z)$, ${\bf{H}}=(H^x,H^y,H^z)$ and ${\bf{B}}=(B^x,B^y,B^z)$ respectively. The vector $\bf{D}$ represents the electric induction $\bf{D}={\varepsilon}\bf{E}$, where ${\varepsilon}$ denotes the dielectric constant of the medium. Using curl on Eq. (6) and the relation that connects magnetization of the medium, finally, we obtain the evolution of the magnetic field component of an EM wave as it propagates through an anisotropic ferromagnetic medium as 
\begin{equation}
\frac{\partial^2}{\partial t^2}({\bf{H}+\bf{M}})=v^2\Bigl[\frac{\partial^2 \bf{H}}{\partial z^2}-\frac{\partial^2 H^z}{\partial z^2}{\bf{\hat{n}}}\Bigr],
\end{equation}
where $v={1\over{\sqrt{\mu\varepsilon}}}$ is the phase velocity of the EM wave in a ferromagnetic medium. Eqs. (2) and (7) are in vector forms of partial differential equations along with the constraint on the magnitude of the spin and solving them analytically is a quite challenging task.
\section{Perturbation scheme and evolution equation}
~~~~~~~We attempt to solve the one-dimensional version of the coupled dynamical Eqs. (2) and (7) using the reductive perturbation method developed by Taniuti and Yajima [33] in order to study the nature of propagation of EM waves in an anisotropic charge-free ferromagnetic medium. The spatial and time variables are stretched in the reductive perturbation method as $\zeta=\epsilon(z-ct)$, $\tau={\epsilon^2}t$ which characterize the shape of the pulse propagating at the speed $c$ and the time variable accounts for the evolution of the propagating pulse, where $\epsilon$ is a small parameter. We rescale $J$ as $\epsilon^{-1}J$, $A$ as ${\epsilon}A$ and $\lambda$ as $\lambda$ on the assumption that the bilinear exchange interaction is stronger than the anisotropic interaction. Additionally, because of the fact that the ferromagnetic medium is anisotropic in nature with an easy axis of magnetization along the $z$ direction, we expand the components of the magnetization $\bf{M}$ of the medium and the magnetic induction $\bf{B}$ in terms of the small parameter $\epsilon$ in a non-uniform way about the uniform values $M_{0}$ and $B_{0}$ respectively by treating the small parameter $\epsilon$ as the perturbation parameter [7] (For more details, refer Appendix-I). By substituting the non-uniform expansions of ${\bf{M}}$ and ${\bf{B}}$ in the component form of Eq. (2) and Eq. (7) and after collecting the coefficients of $\epsilon$ at the order $\epsilon^{0}$, we obtain a system of equations for $\bf{B}_{1}$ and $\bf{M}_{1}$. On solving the resultant equations, we obtain $B_{1}^{x}=k M_{1}^{x}$, $B_{1}^{y}=k M_{1}^{y}$ and $k={B_{0}\over{M_{0}}}={1\over{\varepsilon(v^2-c^2)}}$. In the zeroth order of the perturbation, the steady-state solutions of the $x-$ and $y-$ components of the LLG equation (2) and the Maxwell equation (7) are identically satisfied. We finally obtain the following equations at order $\epsilon^{1}$, by using the results at O($\epsilon^{0}$), as $B_{1}^{z}=0$ and
\begin{eqnarray}
-c{\partial{M_{1}^{x}}\over{\partial\zeta}}=-JM_{0}{\partial^2{M_{1}^{y}}\over{\partial\zeta^2}}-2AM_{0}M_{1}^{y}+{2\beta\over{\mu}}\bigl[B_{1}^{z}M_{1}^{y}-B_{1}^{y}M_{1}^{z}+B_{0}M_{2}^{y}-M_{0}B_{2}^{y}\bigr]\nonumber\\+\lambda\Bigl[J{\partial^2{M_{1}^{x}}\over{\partial\zeta^2}}+2AM_{0}^{2}M_{1}^{x}+{2\beta\over{\mu}}B_{2}^{x}-{2\beta\over{\mu}}\Bigl\{\bigl(B_{1}^{x}M_{1}^{x}+B_{1}^{y}M_{1}^{y}+M_{0}B_{1}^{z}\nonumber\\+B_{0}M_{1}^{z}\bigr)M_{1}^{x}+B_{0}M_{0}M_{2}^{x}\Bigr\}\Bigr],
\end{eqnarray}
\begin{eqnarray}
-c{\partial{M_{1}^{y}}\over{\partial\zeta}}=JM_{0}{\partial^2{M_{1}^{x}}\over{\partial\zeta^2}}+2AM_{0}M_{1}^{x}+{2\beta\over{\mu}}\bigl[B_{1}^{x}M_{1}^{z}-B_{1}^{z}M_{1}^{x}+M_{0}B_{2}^{x}-B_{0}M_{2}^{x}\bigr]\nonumber\\+\lambda\Bigl[J{\partial^2{M_{1}^{y}}\over{\partial\zeta^2}}+2AM_{0}^{2}M_{1}^{y}+{2\beta\over{\mu}}B_{2}^{y}-{2\beta\over{\mu}}\Bigl\{\bigl(B_{1}^{x}M_{1}^{x}+B_{1}^{y}M_{1}^{y}+M_{0}B_{1}^{z}\nonumber\\+B_{0}M_{1}^{z}\bigr)M_{1}^{y}+B_{0}M_{0}M_{2}^{y}\Bigr\}\Bigr],
\end{eqnarray}
and also 
\begin{equation}
{\partial\over{\partial\zeta}}\bigl[B_{2}^{x}-k M_{2}^{x}\bigr]=-2ck\varepsilon{\partial{B_{1}^{x}}\over{\partial\tau}},
\end{equation}
\begin{equation}
{\partial\over{\partial\zeta}}\bigl[B_{2}^{y}-k M_{2}^{y}\bigr]=-2ck\varepsilon{\partial{B_{1}^{y}}\over{\partial\tau}}.
\end{equation}
Making use of Eqs. (10) and (11) in Eqs. (8) and (9), we get
\begin{eqnarray}
-c{\partial{M_{1}^{x}}\over{\partial\zeta}}={4c\varepsilon{\beta}k^{2}M_{0}\over{\mu}}\int_{-\infty}^{\zeta}{{\partial}M_{1}^{y}\over{\partial\tau}}d\zeta-JM_{0}{\partial^{2}M_{1}^{y}\over{\partial\zeta^{2}}}-{2k\beta\over{\mu}}M_{1}^{y}M_{1}^{z}-2AM_{0}M_{1}^{y}\nonumber\\+\lambda\Bigl[J{\partial^2{M_{1}^{x}}\over{\partial\zeta^2}}+2AM_{0}^{2}M_{1}^{x}+{2\beta\over{\mu}}B_{2}^{x}-{2\beta\over{\mu}}\Bigl\{\bigl(B_{1}^{x}M_{1}^{x}+B_{1}^{y}M_{1}^{y}+B_{0}M_{1}^{z}\bigr)M_{1}^{x}\nonumber\\+B_{0}M_{0}M_{2}^{x}\Bigr\}\Bigr]
\end{eqnarray}
and
\begin{eqnarray}
-c{\partial{M_{1}^{y}}\over{\partial\zeta}}=-{4c\varepsilon{\beta}k^{2}M_{0}\over{\mu}}\int_{-\infty}^{\zeta}{{\partial}M_{1}^{x}\over{\partial\tau}}d\zeta+JM_{0}{\partial^{2}M_{1}^{x}\over{\partial\zeta^{2}}}+{2k\beta\over{\mu}}M_{1}^{x}M_{1}^{z}+2AM_{0}M_{1}^{x}\nonumber\\+\lambda\Bigl[J{\partial^2{M_{1}^{y}}\over{\partial\zeta^2}}+2AM_{0}^{2}M_{1}^{y}+{2\beta\over{\mu}}B_{2}^{y}-{2\beta\over{\mu}}\Bigl\{\bigl(B_{1}^{x}M_{1}^{x}+B_{1}^{y}M_{1}^{y}+B_{0}M_{1}^{z}\bigr)M_{1}^{y}\nonumber\\+B_{0}M_{0}M_{2}^{y}\Bigr\}\Bigr].
\end{eqnarray}

To account for the conservation of length of the magnetization $M_{1}^{z}=-{1\over{2M_{0}}}\bigl[(M_{1}^{x})^{2}+(M_{1}^{y})^{2}\bigr]$, we define a new complex field $\psi=M_{1}^{x}-iM_{1}^{y}$ and $|\psi|^2=-2M_{0}M_1^z$ (Refer Appendix-I for more details). The amplitude of the complex field $\psi$ or $|\psi|^2$ corresponds to small deviations from $M_{0}$ (constant, see Eqs. (A2) and (A9)) and phase invariance of $\psi$ corresponds to the isotropy in the $M^{x} - M^{y}$ plane (see Eq. (A8)). By taking a single differentiation of Eqs. (12) and (13) and using the complex field $\psi$ with the appropriate transformations of $Z=\zeta+{A\mu\over{2c{\beta}\varepsilon k^2}}\tau$ and $\tau=\tau$, we obtain the following nonlinear evolution equation after extensive algebra as  
\begin{equation}
i\eta{\partial\psi\over{{\partial}\tau}}+{\partial^2\psi\over{{\partial}Z^2}}+i\sigma{\partial\bigl(|\psi|^2\psi\bigr)\over{{\partial}Z}}-i\delta{\partial^{3}\psi\over{\partial Z^{3}}}+\lambda_{1}{\partial\psi\over{{\partial}\tau}}+\lambda_{2}{\partial^3{\psi}\over{\partial Z^3}}+\lambda_{3}{\partial\bigl(|\psi|^2\psi\bigr)\over{{\partial}Z}}=0,
\end{equation}
where $\eta={4\varepsilon{\beta}k^{2}\over{M_{0}\mu}}$, $\sigma={k{\beta}\over{M_{0}c\mu}}$, $\delta={J\over{M_{0}c}}$, $\lambda_{1}=-{4 \varepsilon{\beta}k^{2}\over{\mu}}\lambda$, $\lambda_{2}={J\over{c}}\lambda$ and $\lambda_{3}=-{k{\beta}\over{c\mu}}\lambda$. After redefinition of Eq. (14), it becomes   
\begin{equation}
\varrho{\partial\psi\over{{\partial}\tau}}+{\partial^2\psi\over{{\partial}Z^2}}+\varphi{\partial^{3}\psi\over{\partial Z^{3}}}+\varsigma{\partial\bigl(|\psi|^2\psi\bigr)\over{{\partial}Z}}=0,
\end{equation}
where $\varrho=i\eta+\lambda_{1}$, $\varphi=-i\delta+\lambda_{2}$ and $\varsigma=i\sigma+\lambda_{3}$. Eq.(15) is known as the time-dependent damped derivative nonlinear Schr\"odinger (TDDNLS) equation. The real part of the first term in Eq. (15) represents the time-dependent damping factor which arisesin the anisotropic ferromagnetic medium when it is subjected to intense EM field in the presence of intrinsic Gilbert damping. This time-dependent damping factor highly influences the nature of nonlinear spin excitations. The second and third terms in Eq. (15) are the dispersion terms that cause the solution $ \psi $ to broaden as it evolves. The third term originates essentially from the exchange interaction between the spins, wherein the real and imaginary components arise due to the presence and absence of Gilbert damping, respectively. The fourth term represents electromagnetically induced higher-order derivative cubic nonlinearity in the material medium. The compensation between the dispersion terms and the electromagnetically induced higher-order derivative cubic nonlinearity admits very interesting nonlinear dynamical structures. The nonlinear evolution Eq. (15) with $\delta=\lambda_{1}=\lambda_{2}=\lambda_{3}=0$ is the completely integrable Kaup-Newell derivative nonlinear Schr\"odinger (DNLS) equation admitting N-soliton solutions [34, 35]. It has been identified in optical fiber that the propagation of optical pulses with asymmetric self-phase modulation and self-steepening is governed by DNLS equation [36]. The nonlinear evolution of Alf\'ven waves that admit completely integrable DNLS equation was first shown by Rogister [37] in space plasma. It is found that the magnetization of an anisotropic ferromagnetic medium is excited in the form of an oscillating electromagnetic soliton as a breathing mode of excitations when it is subjected to a intense EM wave in the absence of Gilbert damping $\lambda=0$ [12]. Therefore, the investigation of nonlinear spin excitations in ferromagnetic media is not only interesting and significant from a mathematical point of view, but also has important physical applications.   
   
\section{Variational methods: Exact soliton solutions}

~~~~~~~The existence of the time-derivative factor in the form of $\partial\psi\over{{\partial}\tau}$ which is proportional to $\lambda_{1}$ in the TDDNLS equation (15) that becomes a disruption to the system causes the parameters associated with the underlying soliton such as amplitude, position, velocity and phase to be time-dependent [38]. The variational method can find the time-dependent parameters associated with the systems under consideration. In order to obtain them through the variational method, the Lagrangian density function $\mathcal{L}$ is constructed by using Eq. (15) as follows: 
\begin{eqnarray}
\mathcal{L}=i\eta\Bigl[\psi^{*}{\partial\psi\over{{\partial}\tau}}-\psi{\partial\psi^{*}\over{{\partial}\tau}}\Bigr]+\Bigl[\psi^{*}{\partial^2\psi\over{{\partial}Z^2}}+\psi{\partial^2\psi^{*}\over{{\partial}Z^2}}\Bigr]+i\sigma\Bigl[\psi^{*}{\partial\bigl(|\psi|^2\psi\bigr)\over{{\partial}Z}}-\psi{\partial\bigl(|\psi|^2\psi^{*}\bigr)\over{{\partial}Z}}\Bigr]\nonumber\\-i\delta\Bigl[\psi^{*}{\partial^{3}\psi\over{\partial Z^{3}}}-\psi{\partial^{3}\psi^{*}\over{\partial Z^{3}}}\Bigr]+\lambda_{1}\Bigl[{\partial\bigl(\psi\psi^{*}\bigr)\over{{\partial}\tau}}\Bigr]+\lambda_{2}\Bigl[\psi^{*}{\partial^3{\psi}\over{\partial Z^3}}+\psi{\partial^3{\psi^{*}}\over{\partial Z^3}}\Bigr]\nonumber\\+\lambda_{3}\Bigl[\psi^{*}{\partial\bigl(|\psi|^2\psi\bigr)\over{{\partial}Z}}+\psi{\partial\bigl(|\psi|^2\psi^{*}\bigr)\over{{\partial}Z}}\Bigr]=0.
\end{eqnarray}
The Lagrangian function $\it{L}$ is obtained by substituting Eq. (16) into the following equation
\begin{equation}
\it{L}=\int_{-\infty}^{\infty}\mathcal{L} dZ.
\end{equation}

\subsection{Gilbert damping managing symmetrically breathing solitons}

~~~~~~~Let us assume that the TDDNLS equation (15) admits a special type of soliton solution that begins to evolve as a function of time with the time-dependent arbitrary functions $\beta(\tau)$, $\alpha(\tau)$ and $\theta(\tau)$ of the form 
\begin{equation}
\psi(Z,\tau)=\beta(\tau)\sech\Bigl[sZ+\alpha(\tau)\Bigr]\exp\Bigl[i\bigl[\theta(\tau)Z+r\bigr]\Bigr].
\end{equation}
After integrating Eq. (17) with respect to $Z$ by making use of Eq. (18), we obtained the following dynamical equation, 
\begin{equation}
{d\beta\over{d\tau}}={1\over{3\lambda_{1}}}\Bigl[s^2\beta+3{\delta}s^2\theta\beta+2\sigma\theta\beta^3+3\beta\theta^2+3{\delta}\beta\theta^3\Bigr].
\end{equation}
Next, we multiply both sides of Eq. (17) by $\tanh(sZ+\alpha(\tau))$ and integrate it once to obtain
\begin{equation}
5s\lambda_{1}{d\alpha\over{d\tau}}-15\eta{d\theta\over{d\tau}}=7\lambda_{2}s^4-6\lambda_{3}s^2\beta^2+15\lambda_{2}s^2\theta^2. 
\end{equation}
After rearranging the above Eq. (20), we get the following equation in which  both LHS and RHS are equal if they evolve independently at a constant rate $m$, and hence
\begin{equation}
15\eta{d\theta\over{d\tau}}+15\lambda_{2}s^2\theta^2=5s\lambda_{1}{d\alpha\over{d\tau}}-7\lambda_{2}s^4+6\lambda_{3}s^2\beta^2=m. 
\end{equation}
The LHS and RHS of the aforementioned equation now become
\begin{equation}
15\eta{d\theta\over{d\tau}}+15\lambda_{2}s^2\theta^2=m 
\end{equation}
and
\begin{equation}
5s\lambda_{1}{d\alpha\over{d\tau}}-7\lambda_{2}s^4+6\lambda_{3}s^2\beta^2=m.                
\end{equation}
If the time-dependent function $\beta(\tau)$ remains constant ($\beta(\tau)=\beta$), the following solutions can be obtained by solving Eqs. (22 $\&$ 23),
\begin{subequations}
\begin{align}
\alpha(\tau)={1\over{5s\lambda_{1}}}\Bigl[{9\delta\lambda_{3}s^4\over{\sigma}}+7\lambda_{2}s^4+m\Bigr]\tau+c_{2},\\ 
\theta(\tau) = {1\over{s}}\sqrt{{m\over{15\lambda_{2}}}}\tanh\Biggl[{s\over{\eta}}\sqrt{{m\lambda_{2}\over{15}}}\bigl(\tau+c_{1}\bigr)\Biggr],\\
\beta(\tau)=\beta=\pm{s\over{2}}\sqrt{{-6\delta\over{\sigma}}}.               
\end{align}
\end{subequations}
where $c_{1}$ and $c_{2}$ are constants of integration. Using the complex field $\psi=M_{1}^{x}-iM_{1}^{y}$, $|\psi|^2=-2M_{0}M_1^z$ and Eqs. (24), the components $M_{1}^{x}$, $M_{1}^{y}$ and $M_{1}^{z}$ of the magnetization of the anisotropic ferromagnetic material exposed by intense EM field can be constructed as
\begin{subequations}
\begin{align}
M_{1}^{x}=\pm{s\over{2}}\sqrt{{-6\delta\over{\sigma}}}\sech\Biggl[sZ+{1\over{5s\lambda_{1}}}\Bigl[{9\delta\lambda_{3}s^4\over{\sigma}}+7\lambda_{2}s^4+m\Bigr]\tau+c_{2}\Biggr]\cos\Bigl[\theta(\tau)Z+r\Bigr],\\
M_{1}^{y}=\mp{s\over{2}}\sqrt{{-6\delta\over{\sigma}}}\sech\Biggl[sZ+{1\over{5s\lambda_{1}}}\Bigl[{9\delta\lambda_{3}s^4\over{\sigma}}+7\lambda_{2}s^4+m\Bigr]\tau+c_{2}\Biggr]\sin\Bigl[\theta(\tau)Z+r\Bigr],\\
M_{1}^{z}=-{u(Z,\tau)^2\over{2M_{0}}}={3s^2\delta\over{4{\sigma}M_{0}}}\sech^{2}\Biggl[sZ+{1\over{5s\lambda_{1}}}\Bigl[{9\delta\lambda_{3}s^4\over{\sigma}}+7\lambda_{2}s^4+m\Bigr]\tau+c_{2}\Biggr].
\end{align}
\end{subequations}

When $\lambda=0$, the propagation of intensely varying EM wave along the direction of easy axis of an anisotropic ferromagnetic medium ($M^{z}_{1}$) is governed by a highly localized magnetized state in the form of soliton which makes some instabilities in the transmission character of the $M^{x}_{1}$ and $M^{y}_{1}$ components [12]. Therefore, higher-order local instabilities arise in the form of derivative cubic nonlinearity in the medium which may result from the periodically varying refractive index of the medium. As a result, the $M^{x}_{1}$ and $M^{y}_{1}$ components of magnetization of the medium exhibit oscillating solitary modes in the form of breathers with constant amplitude due to the strong counter-balance between the higher-order dispersion and the electromagnetically induced derivative cubic nonlinearity. In order to realize the influence of Gilbert damping with third-order dispersion and derivative cubic nonlinearity when intense varying EM wave propagates through an anisotropic ferromagnetic medium, we have plotted Eqs. (25) for the x, y and z-components of the magnetization $M^{x}_{1}$, $M^{y}_{1}$ and $M^{z}_{1}$ with parametric choice of $\eta=-0.5$, $\sigma=-0.3$, $\delta=0.3$, $M_{0}=0.153$, r=0.2, s=-0.1, $C_{1}= 0.2$, $C_{2}= 0.01$, $m = 0.04$ for various values of $\lambda$.

\begin{figure}[!h]
	\centering
	(a)\includegraphics[width=0.28\linewidth]{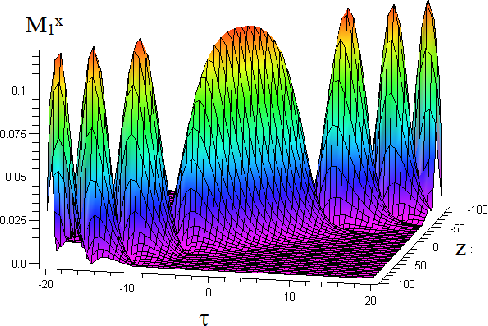}
	(b)\includegraphics[width=0.28\linewidth]{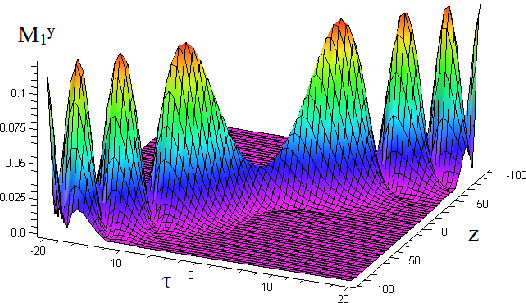}
	(c)\includegraphics[width=0.28\linewidth]{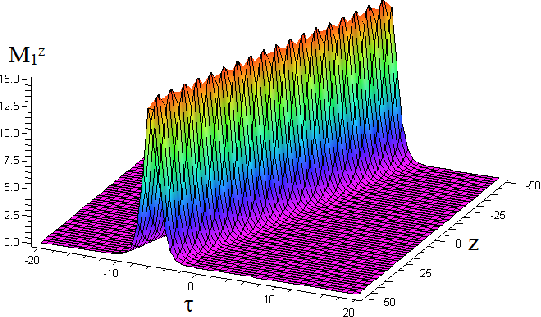}
	\caption{Evolution of symmetrically breathing solitons in (a) $M_{1}^{x}$ and (b) $M_{1}^{y}$, and soliton in (c) $M_{1}^{z}$ for $\eta=-0.5$, $\sigma=-0.3$, $\delta=0.3$, $M_{0}=0.153$, r=0.2, s=-0.1, $C_{1}= 0.2$, $C_{2}= 0.01$, m = 0.04 and $\lambda=0.03$.}
\end{figure}

\begin{figure}[!h]
	\centering
	(a)\includegraphics[width=0.28\linewidth]{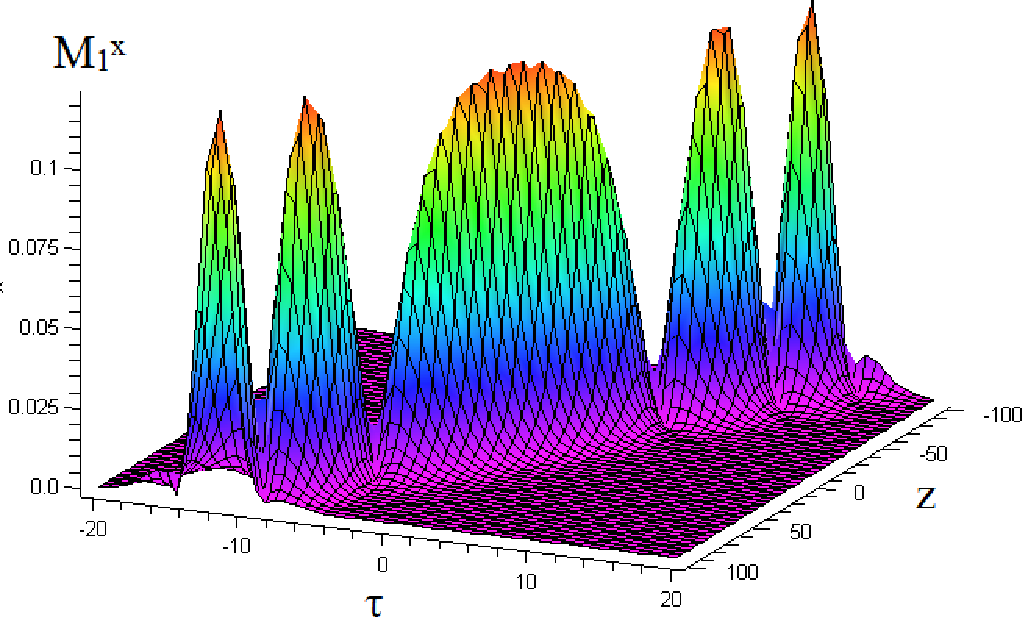}
	(b)\includegraphics[width=0.28\linewidth]{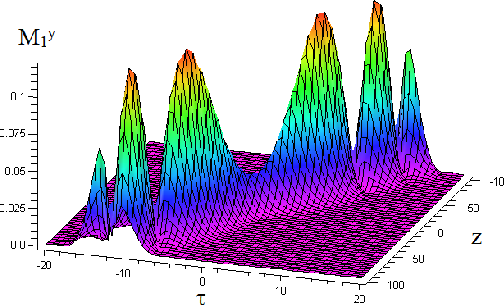}
	(c)\includegraphics[width=0.28\linewidth]{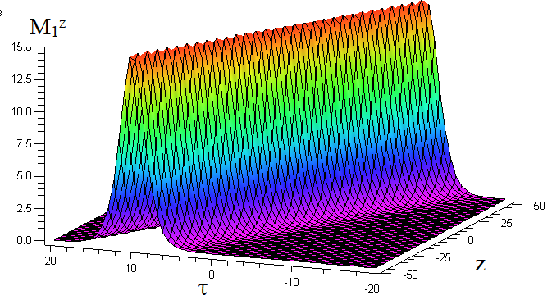}
	\caption{Evolution of symmetrically breathing solitons in (a) $M_{1}^{x}$ and (b) $M_{1}^{y}$, and soliton in (c) $M_{1}^{z}$ for $\eta=-0.5$, $\sigma=-0.3$, $\delta=0.3$, $M_{0}=0.153$, r=0.2, s=-0.1, $C_{1}= 0.2$, $C_{2}= 0.01$, m = 0.04 and $\lambda=0.05$.}
\end{figure}

\begin{figure}[!h]
	\centering
	(a)\includegraphics[width=0.28\linewidth]{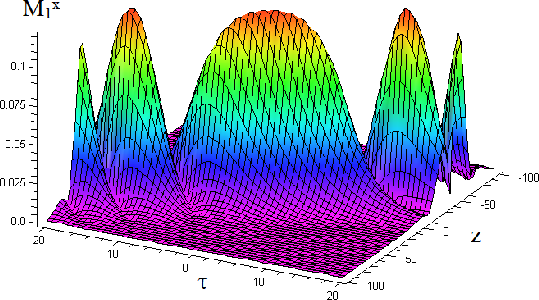}
	(b)\includegraphics[width=0.28\linewidth]{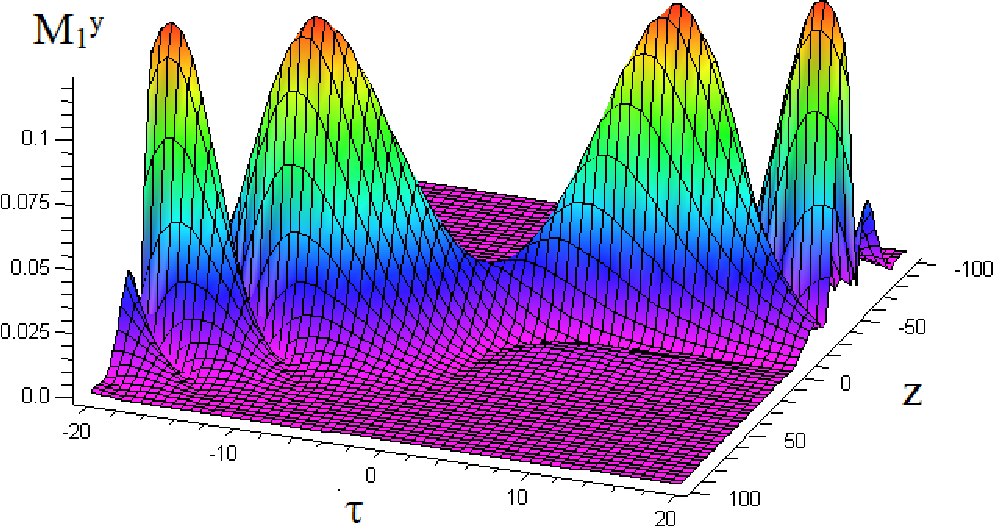}
	(c)\includegraphics[width=0.28\linewidth]{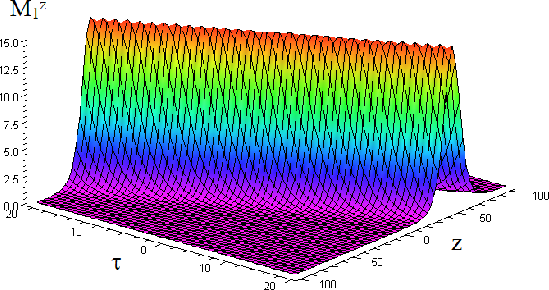}
	\caption{Evolution of symmetrically breathing solitons in (a) $M_{1}^{x}$ and (b) $M_{1}^{y}$, and soliton in (c) $M_{1}^{z}$ for $\eta=-0.5$, $\sigma=-0.3$, $\delta=0.3$, $M_{0}=0.153$, r=0.2, s=-0.1, $C_{1}= 0.2$, $C_{2}= 0.01$, m = 0.04 and $\lambda=0.1$.}
\end{figure}

When an anisotropic ferromagnetic medium is exposed to a strong EM field with $\lambda=0.03$, the z-component of magnetization $M^{z}_{1}$ experiences nonlinear spin excitations in the form of soliton, as illustrated in Fig. 1c. This soliton spin excitations in $M^{z}_{1}$ supports Gilbert damping managing symmetrically oscillating solitary modes in the form of breathing soliton. It has been found that the x-component of magnetization $M^{x}_{1}$ of the ferromagnetic medium experiences nonlinear spin excitations in the form of symmetrically breathing soliton in the presence of intense EM field and Gilbert damping as seen in Fig. 1a. That is, in the initial stage of breathing soliton, the width of the central peak of breathing soliton is maximum. As time goes on, the Gilbert damping causes the width of the breathing soliton to slightly decrease while maintaining its amplitude. This breathing soliton absorbs the electromagnetic energy from the intense EM field during the propagation in order to manage the influence of inherent Gilbert damping present in the system which results in lossless amplitude of breathing soliton. Also the y-component of magnetization $M^{y}_{1}$ of the ferromagnetic medium exhibits the same behaviors without the central maxima peak as depicted in Fig. 1b. This is due to the fact that the counterbalance between the electromagnetically induced derivative cubic nonlinearity and time-dependent damping factor ${\partial\psi\over{{\partial}\tau}}$ in the anisotropic ferromagnetic medium having intrinsic Gilbert damping leads to symmetrically breathing solitons with varying width. On the other hand, it has been found that, in the absence of Gilbert damping, electromagnetically induced derivative cubic nonlinearity in an anisotropic ferromagnetic medium admits breathing soliton with constant amplitude and width due to intense EM field [12]. However, an isotropic Heisenberg ferromagnetic spin chain with Gilbert damping in the absence intense EM field is governed by a damped nonlinear Schr\"odinger equation which admits decaying solitary wave solution for the energy density and magnetization density [39]. When natural Gilbert damping $\lambda$ is 0.05, the z-component of magnetization shows the same behavior as described previously for $\lambda= 0.03$ with small change in phase shift. Meanwhile, the x and y-components of magnetization exhibit low breathing frequency of symmetrically breathing solitons with a small change in phase shift as shown in Figs. 2. As the value of Gilbert damping is increased to $\lambda=0.1$, the x, y and z components of magnetization undergo nonlinear spin excitations with small changes in phase shifts as well as breathing frequencies as depicted in Figs. 3. From the observations, it is inferred that when the value of Gilbert damping is increased in the presence of intense EM field, both the x and y components of magnetization of the anisotropic ferromagnetic medium show Gilbert damping managing symmetrically breathing solitons. Thus, the Gilbert damping managing symmetrically breathing soliton could open new opportunities for robust information processing technologies.

\subsection{Breathing dromions-like spin excitations}
~~~~~~~Next, we assume that the solution with the time-dependent arbitrary function $\alpha(\tau)$ as
\begin{equation}
\psi(Z,\tau)=\alpha(\tau)\sech\Bigl[sZ+\alpha(\tau)\Bigr]\exp\Bigl[i\bigl[\alpha(\tau)Z+r\bigr]\Bigr],
\end{equation}
where, $r$ and $s$ represent arbitrary constants. Note that the form (26) is a special case of (18). Allowing the time-dependent parameter $\alpha(\tau)$ to evolve as a function of time is necessary to understand the nature of spin excitations associated with the TDDNLS equation. The following evolution equation can be obtained by using Eq. (26) and multiplying $\tanh(sZ+\alpha(\tau))$ on both sides of Eq. (17), and then integrating it with regard to $Z$ as
\begin{equation}
15\eta{d\alpha\over{d\tau}}-10s\lambda_{1}{d\alpha\over{d\tau}}+14s^4\lambda_{2}-12s^2\lambda_{3}\alpha^2+30s^2\lambda_{2}\alpha^2=0.
\end{equation}
By solving the nonlinear ordinary differential Eq. (27), the following solution can be obtained:
\begin{equation}
\alpha(\tau) = A_{p}\tanh\bigl[\Omega(\tau+C)\bigr],
\end{equation}
where $A_{p}={s\sqrt{(42\lambda_{2}\lambda_{3}-105\lambda_{2}^2})\over{(15\lambda_{2}-6\lambda{3})}}$, $\Omega={2\over{5}}s^3{\sqrt{(42\lambda_{2}\lambda_{3}-105\lambda_{2}^2})\over{(3\eta-2s\lambda_{1})}}$ and $C$ represents integration constant. The evolution of magnetization components of the anisotropic ferromagnetic medium when subjected to an EM field can be constructed using the solution Eq. (28) in the soliton solution (26) through the complex field $\psi$ as 
\begin{subequations}
\begin{align}
M_{1}^{x}=A_{p}\tanh\bigl[\Omega(\tau+C)\bigr]\sech\Bigl[sZ+A_{p}\tanh\bigl[\Omega(\tau+C)\bigr]\Bigr]\cos\Bigl[A_{p}\tanh\bigl[\Omega(\tau+C)\bigr]Z+r\Bigr],\\
M_{1}^{y}=-A_{p}\tanh\bigl[\Omega(\tau+C)\bigr]\sech\Bigl[sZ+A_{p}\tanh\bigl[\Omega(\tau+C)\bigr]\Bigr]\sin\Bigl[A_{p}\tanh\bigl[\Omega(\tau+C)\bigr]Z+r\Bigr],\\
M_{1}^{z}=-{u(Z,\tau)^2\over{2M_{0}}}=-{A_{p}^2\over{2M_{0}}}\Biggl[\tanh^2\bigl[\Omega(\tau+C)\bigr]\sech^2\Bigl[sZ+A_{p}\tanh\bigl[\Omega(\tau+C)\bigr]\Bigr]\Biggr].
\end{align}
\end{subequations}

\begin{figure}[!h]
	\centering
	(a)\includegraphics[width=0.29\linewidth]{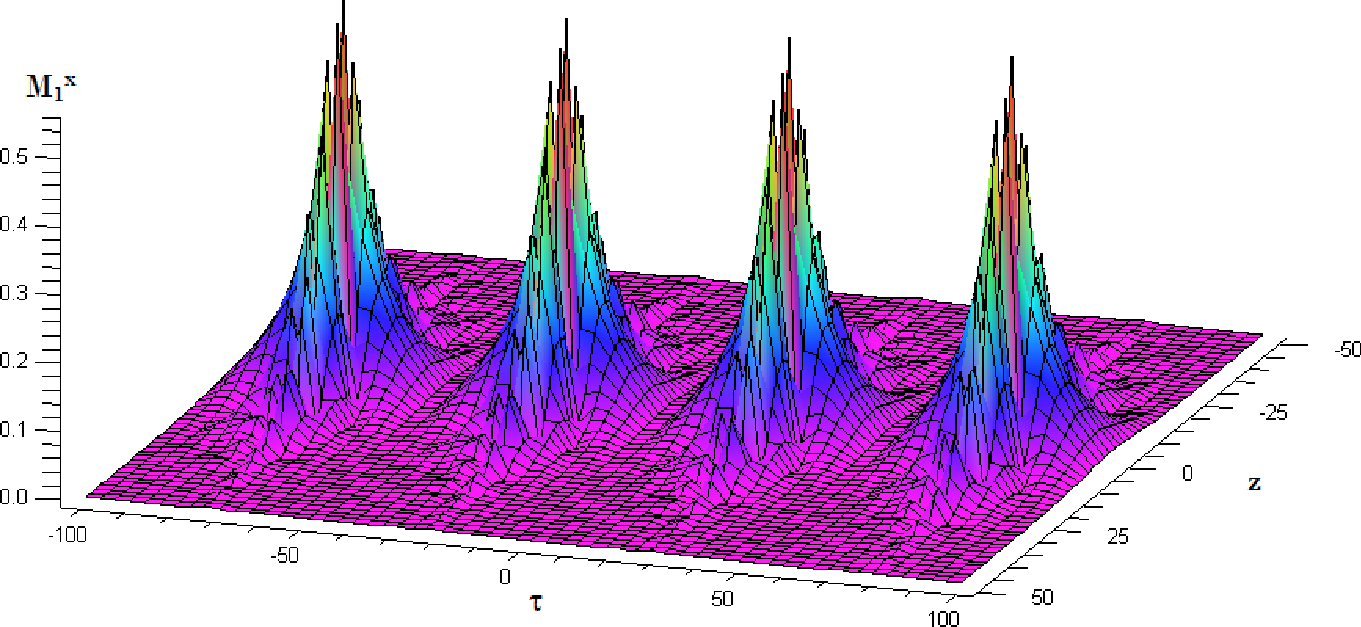}
	(b)\includegraphics[width=0.29\linewidth]{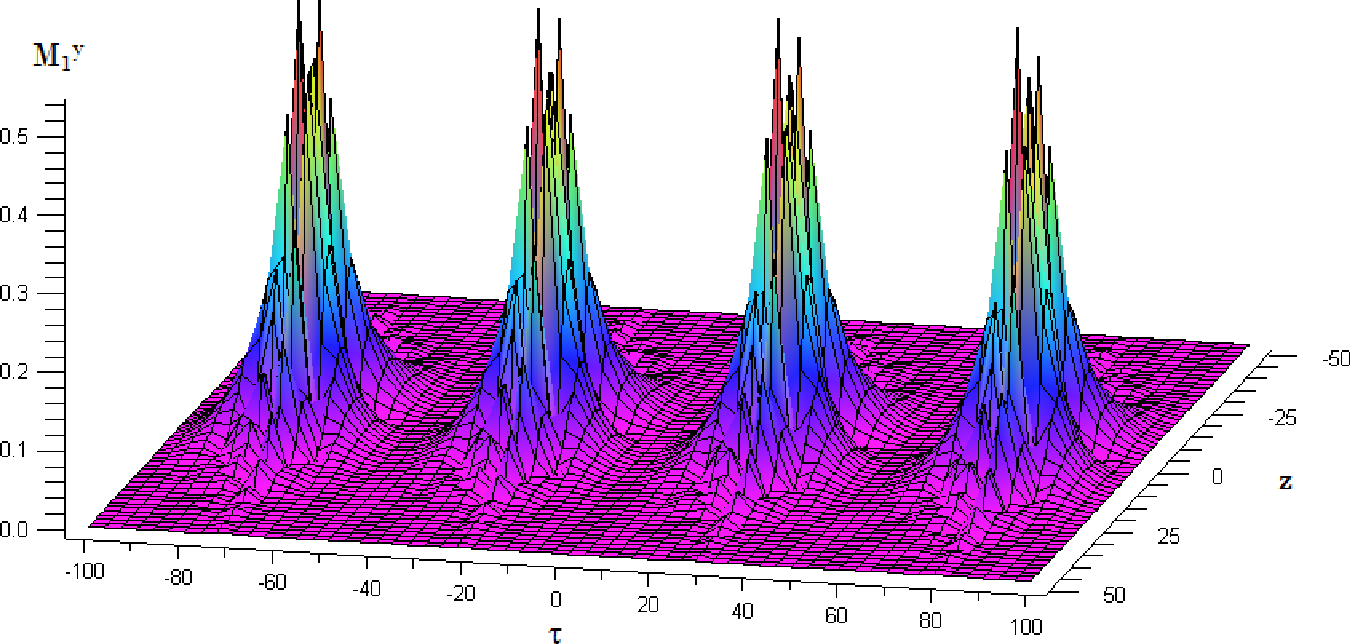}
	(c)\includegraphics[width=0.29\linewidth]{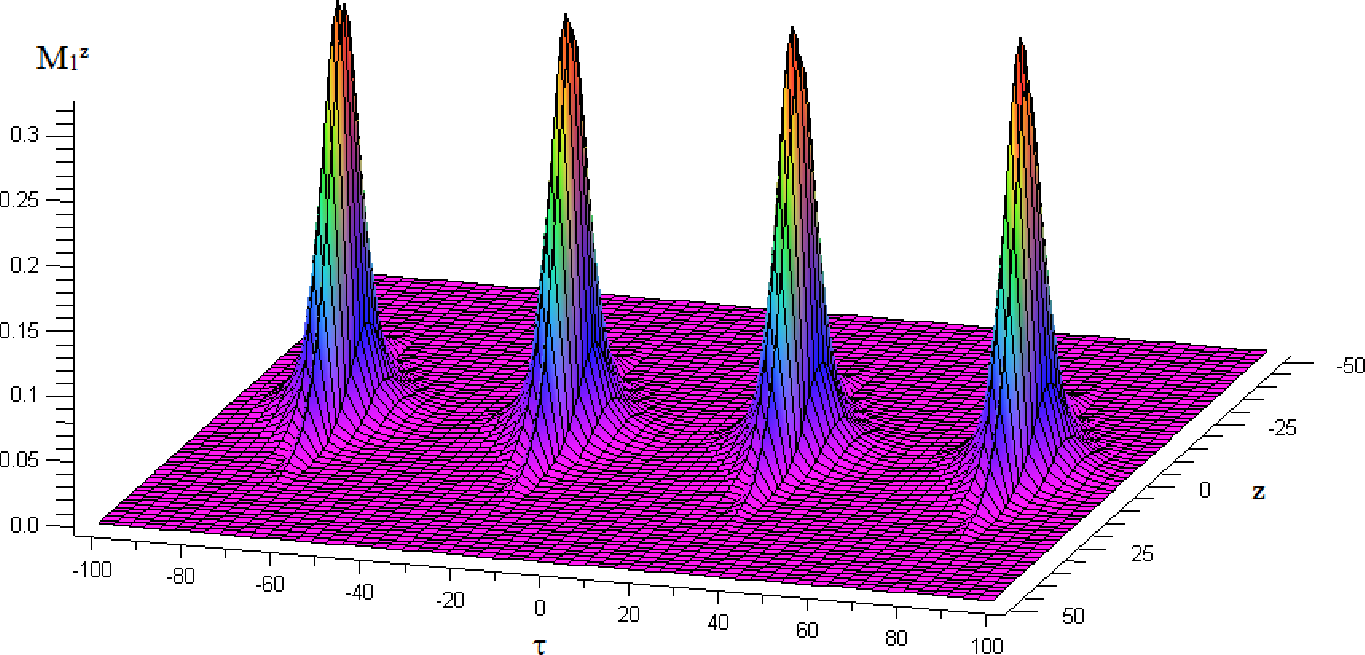}
	\caption{Evolution of localized erupting electromagnetic (a) $M_{1}^{x}$ and (b) $M_{1}^{y}$ breathing dromion-like modes of excitations, and (c) $M_{1}^{z}$ breathing dromion-like soliton for $\eta=0.01$, $\sigma=0.2$, $\delta=0.31$, $M_{0}=0.1$, r=0.1, s=0.13 and $\lambda=0.05$.}
\end{figure}

The x, y and z-components of the magnetization $M^{x}_{1}$, $M^{y}_{1}$ and $M^{z}_{1}$ have been plotted using Eqs. (29) to understand the nature of nonlinear spin excitations with a parametric choice of $\eta=0.01$, $\sigma=0.2$, $\delta=0.31$, $M_{0}=0.1$, $r=0.1$ and $s=0.13$ for the different values of $\lambda$. The propagation of varying EM wave through an anisotropic ferromagnetic medium in the presence of Gilbert damping is governed by TDDNLS equation in which the time-dependent damping factor ${\partial\psi\over{{\partial}\tau}}$ disrupts the periodically varying refractive index of the medium. More interestingly, it is found that when $\lambda=0.05$, spatially localized and temporally periodic dromion-like electromagnetic soliton is observed in the $M^{z}_{1}$ component of the anisotropic ferromagnetic system as shown in Fig. 4c. Generally, the dromions decay exponentially and are localized in all the spatial directions which is a special type of higher dimensional soliton solutions. Here, the $M^{x}_{1}$ and $M^{y}_{1}$ components of magnetization experiences oscillating electromagnetic dromion-like modes of excitations with erupting profiles as shown in Figs. 4a $\&$ 4b. The evolution of erupting soliton starts from a localized stationary solution which has a perfect dromion-like shape. After a while its profiles become covered with small ripples due to internal instability which seem to move downwards and very soon the pulse is covered with this seemingly fluctuating structure. As the ripples grow in size, the oscillating electromagnetic dromion splits into fragments, analogous to a mountain after a powerful volcanic eruption or an earthquake. This fluctuating structure, but well-localized, repeats itself exactly in successive periods with constant amplitude. As a result of the nonlinear Gilbert damping effect in the presence of EM wave interaction with spins, soliton eruptions or explosions may occur during the transition from one stable spin configuration to another spin configuration.

\begin{figure}[!h]
	\centering
	(a)\includegraphics[width=0.29\linewidth]{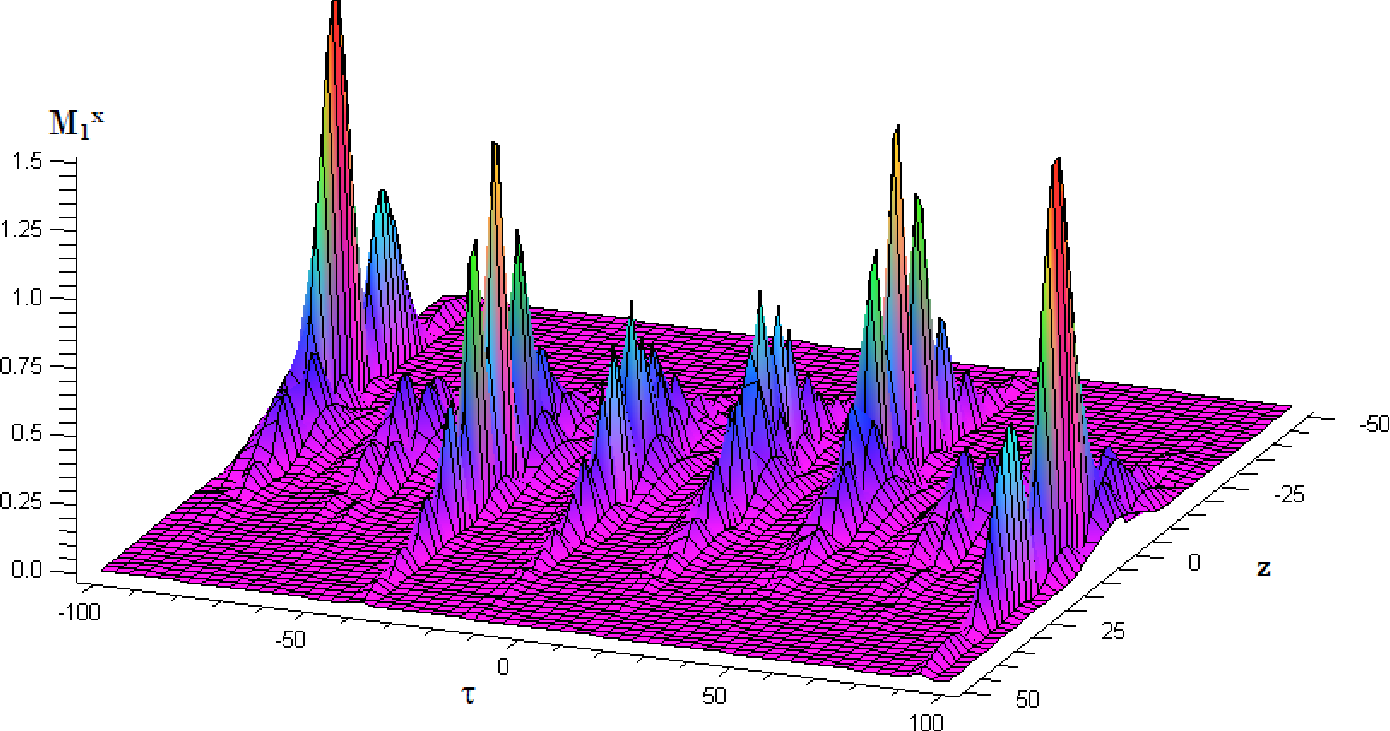}
	(b)\includegraphics[width=0.29\linewidth]{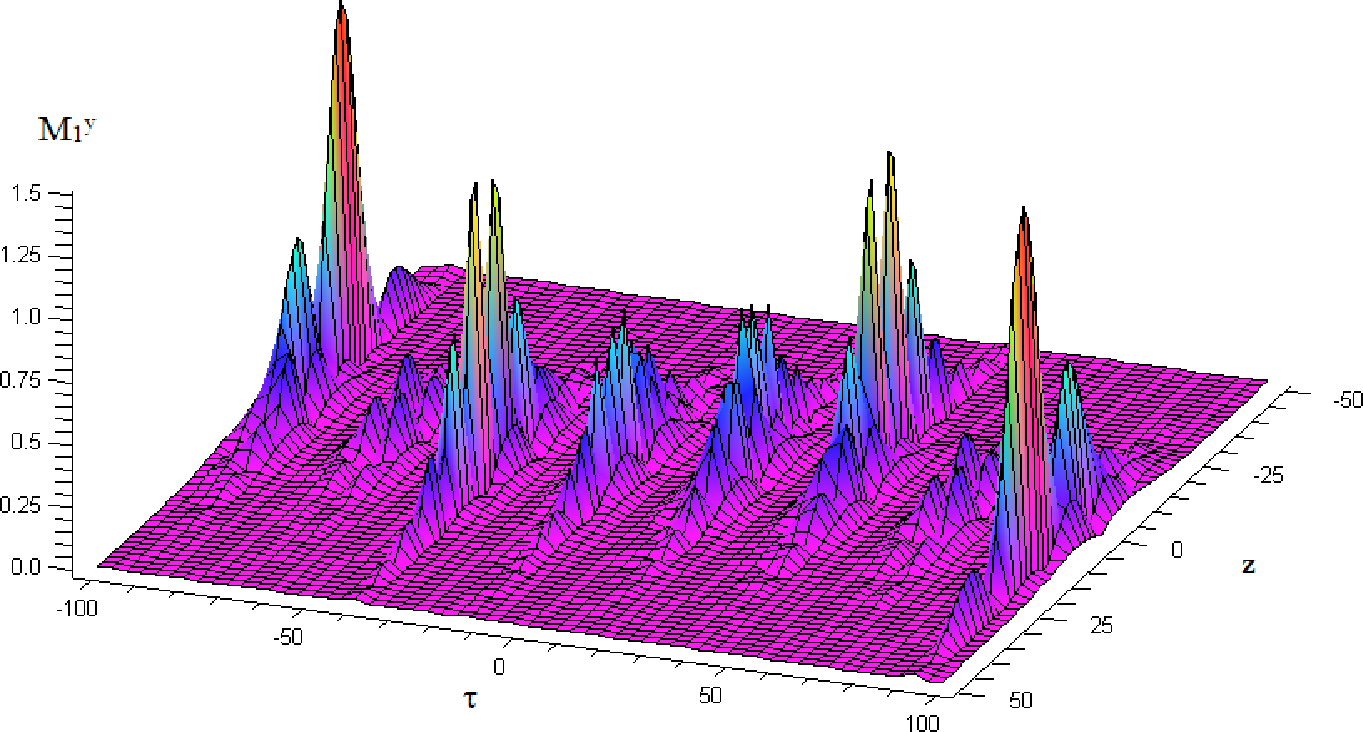}
	(c)\includegraphics[width=0.29\linewidth]{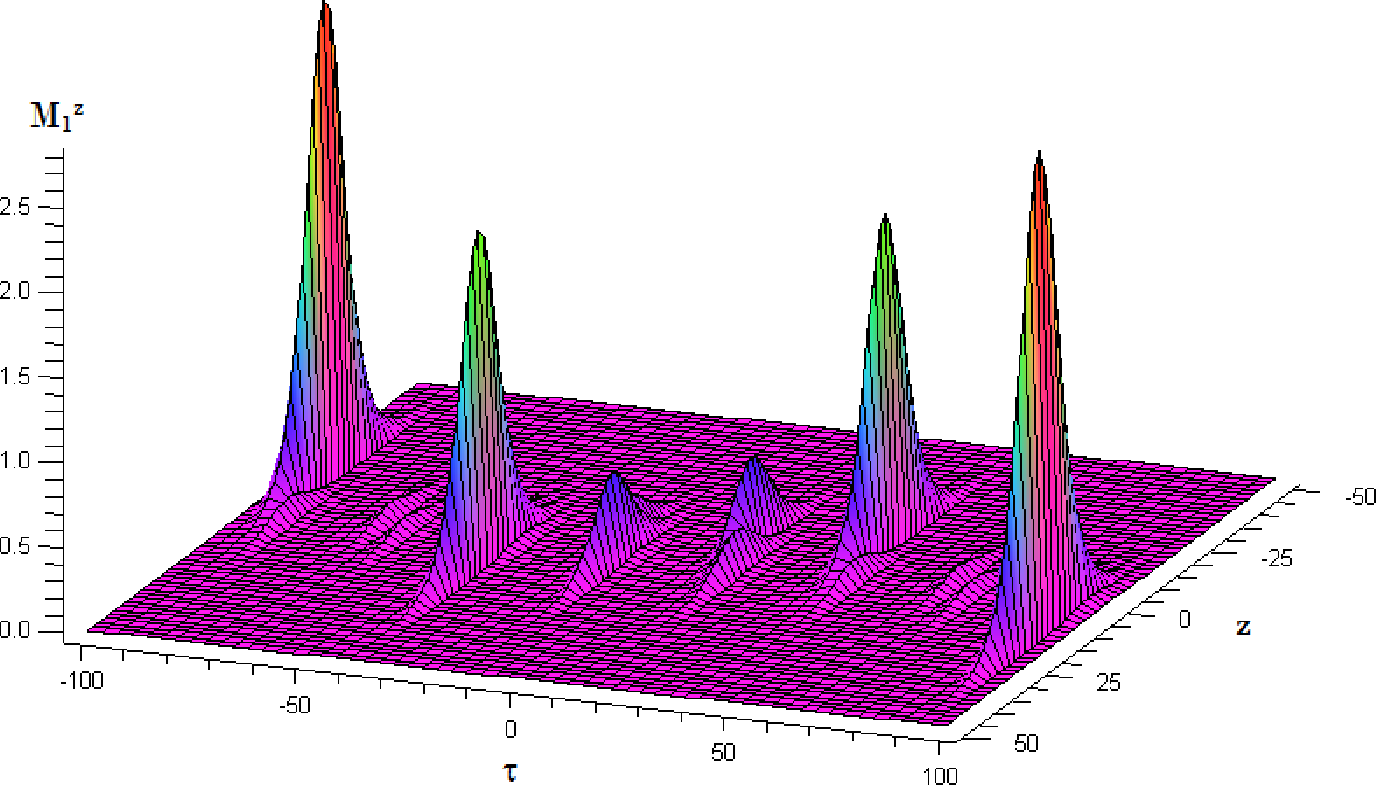}
	\caption{Evolution of localized erupting electromagnetic (a) $M_{1}^{x}$ and (b) $M_{1}^{y}$ breathing dromion-like modes of excitations, and (c) $M_{1}^{z}$ breathing dromion-like soliton with varying amplitudes for $\eta=0.01$, $\sigma=0.2$, $\delta=0.31$, $M_{0}=0.1$, r=0.1, s=0.13 and $\lambda=0.1$.}
\end{figure}

\begin{figure}[!h]
	\centering
	(a)\includegraphics[width=0.29\linewidth]{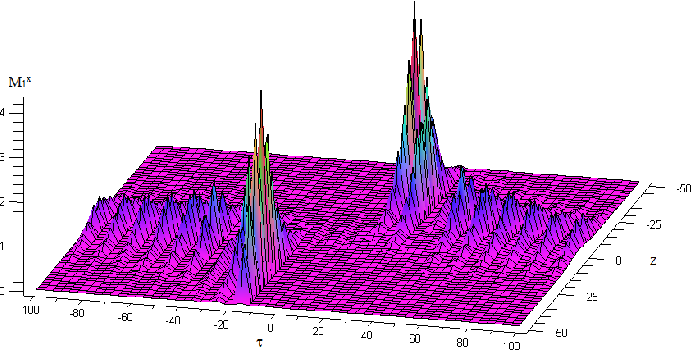}
	(b)\includegraphics[width=0.29\linewidth]{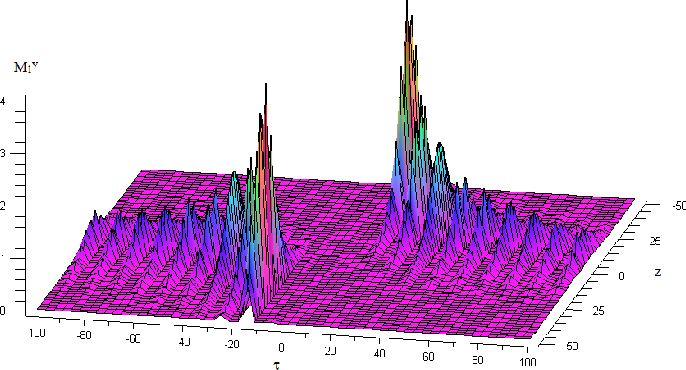}
	(c)\includegraphics[width=0.29\linewidth]{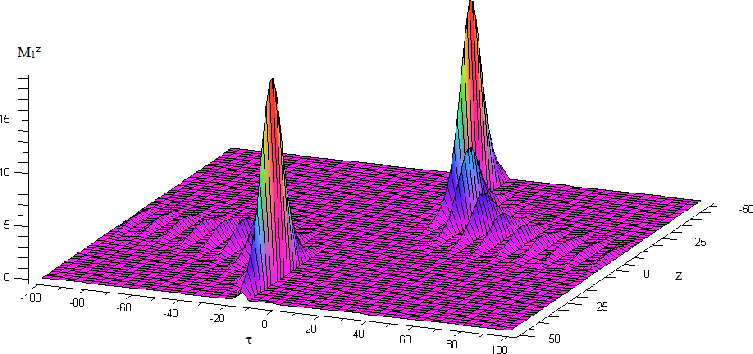}
	\caption{Evolution of localized erupting electromagnetic (a) $M_{1}^{x}$ and (b) $M_{1}^{y}$ breathing dromion-like modes of excitations, and (c) $M_{1}^{z}$ breathing dromion-like soliton with decaying amplitudes for $\eta=0.01$, $\sigma=0.2$, $\delta=0.31$, $M_{0}=0.1$, r=0.1, s=0.13 and $\lambda=0.3$.}
\end{figure}
\begin{figure}[!h]
	\centering
	(a)\includegraphics[width=0.29\linewidth]{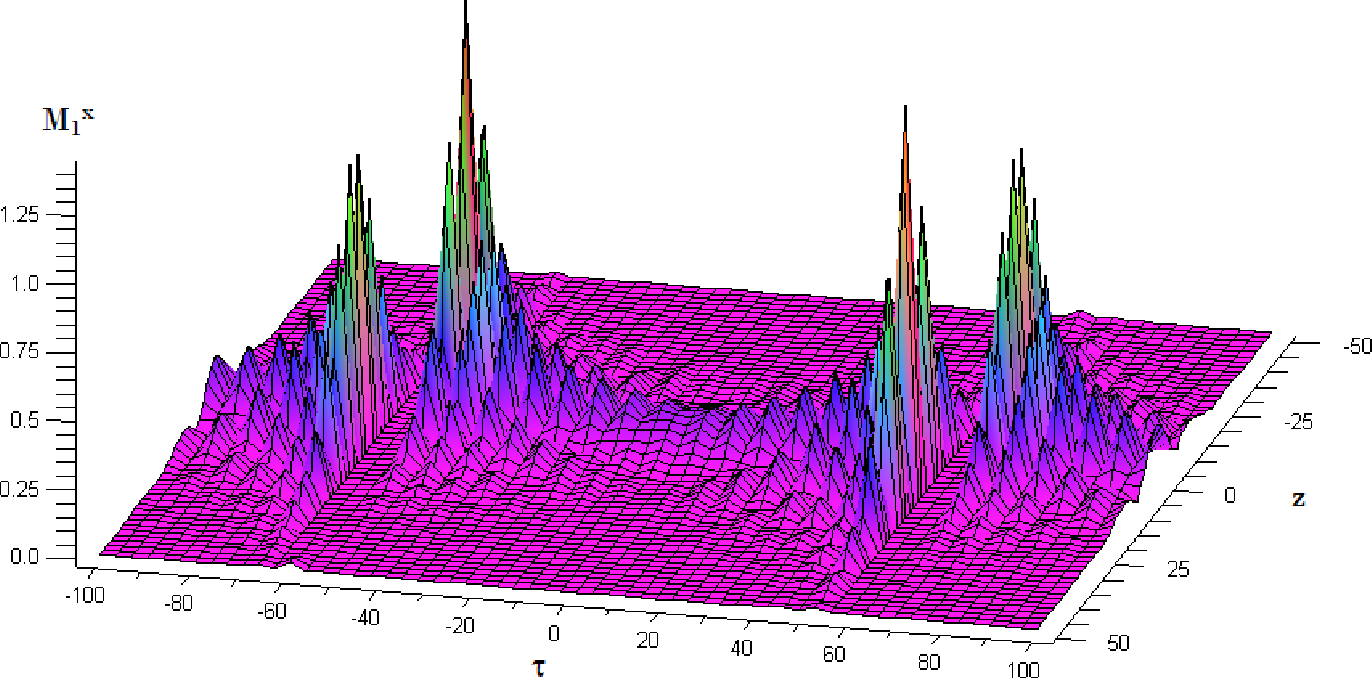}
	(b)\includegraphics[width=0.29\linewidth]{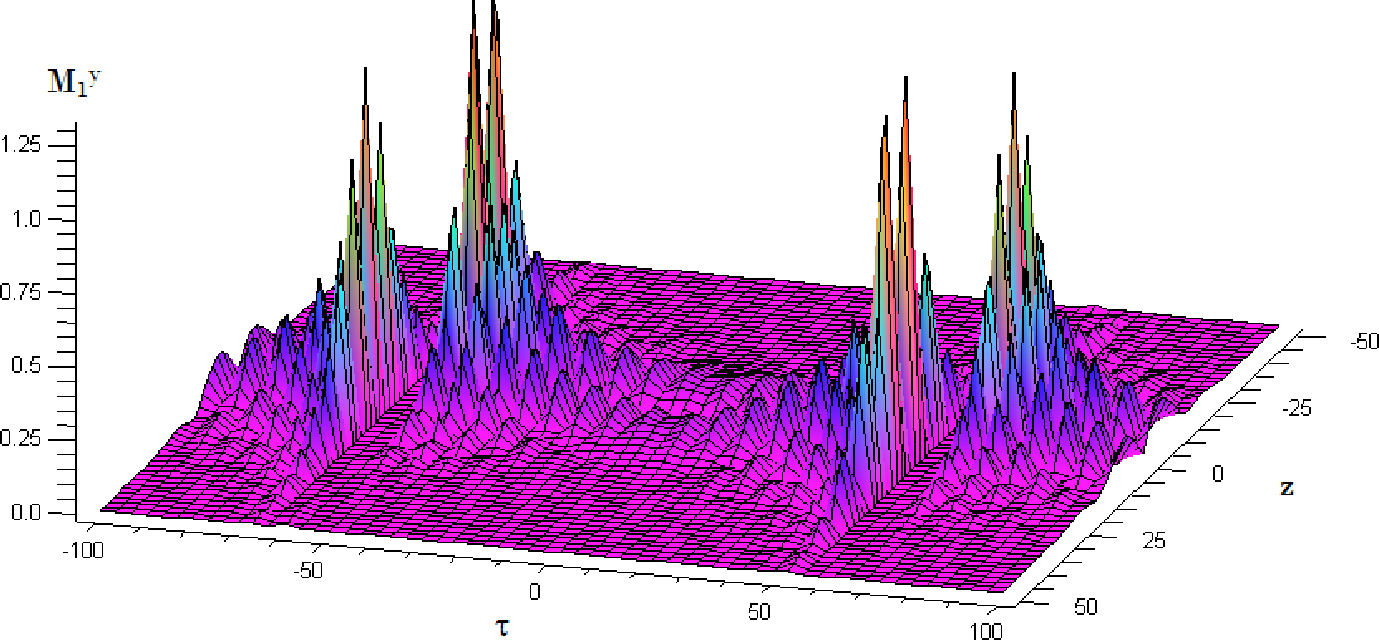}
	(c)\includegraphics[width=0.29\linewidth]{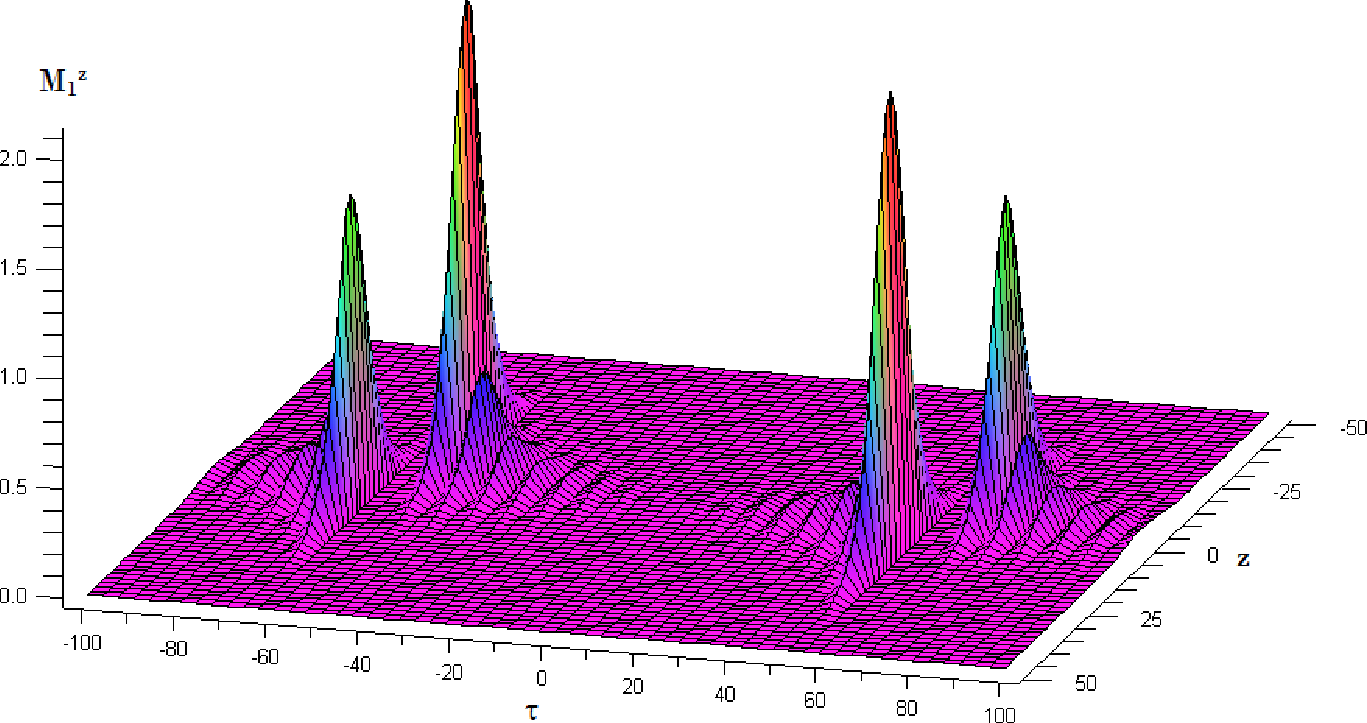}
	\caption{Evolution of localized erupting electromagnetic (a) $M_{1}^{x}$ and (b) $M_{1}^{y}$ breathing dromion-like modes of excitations, and (c) $M_{1}^{z}$ breathing dromion-like modes of excitations with growing-decaying amplitudes for $\eta=0.01$, $\sigma=0.2$, $\delta=0.31$, $M_{0}=0.1$, r=0.1, s=0.13 and $\lambda=0.5$.}
\end{figure}

When $\lambda=0.1$ with intense EM wave, the amplitude of the oscillating electromagnetic dromions is suppressed in the $M^{z}_{1}$ component and become localized solitary modes with varying amplitude, while the $M^{x}_{1}$ and $M^{y}_{1}$ components of magnetization appear as erupting dromion-like modes with varying amplitude as shown in Figs. 5.  When $\lambda=0.3$, the amplitude of the localized solitary modes of the $M^{z}_{1}$ components of magnetization decreases exponentially as the time goes. However, the $M^{x}_{1}$ and $M^{y}_{1}$ components of magnetization exhibit erupting solitary excitations with decaying amplitude as shown in Figs. 6. From the figures, it is inferred that when the strength of the Gilbert damping is high in the presence of EM wave, the precessional motion of the spin vector experiences strong nonlinear force along the z-direction in the anisotropic ferromagnetic medium which causes instability in the x-y plane. This strong nonlinear force may be developed physically from the imperfections, impurity atoms, or defects in the regular lattice sites of the anisotropic ferromagnetic medium, which leads to the instability in the precessional motion of the spin vector. A fluctuating or nonlinear precessional motion of the spin vector due to the instability will be relaxed along the z-direction resulting in the decaying amplitude of dromion-like modes of excitations with erupting profiles in the $M^{x}_{1}$ and $M^{y}_{1}$ components of magnetization. 

When the Gilbert damping is increased to $\lambda=0.5$ in the presence of an EM wave, the amplitude of the nonlinear excitations in the $M^{z}_{1}$ components of magnetization progressively grows to a maximum value and then subsequently diminishes to a minimum value. Here, the energy of the nonlinear spin excitations increases gradually to create dromion-like structure until it reaches a maximum level and it experiences intermittency at that point and then it begins to dissipate its energy as decaying or annihilating dromions. It is clearly seen in the $M^{z}_{1}$ components of magnetization as in Figs .7. The $M^{x}_{1}$ and $M^{y}_{1}$ components of magnetization also exhibit the same behavior with erupting profiles. To preserve their originality, the dromion-like modes of excitation restore themselves after each eruption. Therefore, when an anisotropic ferromagnetic medium is subjected to a intense EM wave in the presence of a strong dissipation in the form of Gilbert damping, an unexpected event of creation and annihilation of nonlinear spin excitations in the form of growing and decaying dromion-like mode is observed in it which may also offer potentially attractive platforms for robust information processing. We also wish to finally state the following limitation of our study. Even though our study brings out the existence of a very interesting localized dromion-like structure, we have not established the stability of these structures. We hope to pursue such studies in the near future.
\section{Conclusions}

~~~~~~~The propagation of EM waves in an anisotropic ferromagnetic medium under the influence of Gilbert damping has been studied theoretically. The interaction of the magnetic field component of the intense EM wave with the magnetization of a ferromagnetic medium has been studied by mapping the associated Maxwell's equations coupled with a LLG equation. When the magnetization of the ferromagnetic medium and magnetic field component of the EM wave are perturbed in a non-uniform way by using the reductive perturbation method, the associated nonlinear spin excitations are governed by a time-dependent damped derivative nonlinear Schr\"odinger equation. The Lagrangian density function has been constructed by using the variational method to understand the nature of nonlinear spin excitations under the influence of the time-dependent damping on the system under consideration. It has been demonstrated  for the solution (18) that  Gilbert damping managing symmetrically breathing solitons is admitted by the compensation between the electromagnetically induced derivative cubic nonlinearity and the time-dependent damping factor in an anisotropic ferromagnetic medium. In particular, for the solution (26), it has been shown that the z-component of magnetization of the medium is governed by spatially localized and temporally periodic electromagnetic breathing dromion-like soliton, while the x- and y-components of the medium are excited in the form of erupting electromagnetic breathing dromion-like modes of excitations when the strength of Gilbert damping is low. When the strength of Gilbert damping is high, it is found that the propagation of EMW in a ferromagnetic medium is governed by decaying breathing dromion-like modes of excitations and an unexpected creation-annihilation modes of excitations in the form of growing-decaying dromion-like modes. Therefore, it is theoretically understood from the present study that the propagation of EM wave in an anisotropic ferromagnetic medium with Gilbert damping admits very interesting nonlinear dynamical structures and phenomenon which may have potentially attractive platforms for robust information processing.  

\section*{Acknowledgment}
The author M. Lakshmanan wishes to thank the Science and Engineering Research Board, Department of Science and Technology, Government of India for the award of a National Science Chair position under Grant No. NSC/2020/00029.
\appendix

\section*{Appendix-I}

The non-uniform expansion of the components of the magnetization ${\bf{M}}$ of the medium and the magnetic induction ${\bf{B}}$ in terms of the small parameter ${\epsilon}$ as the perturbation parameter about the uniform values $M_{0}$ and $B_{0}$ respectively as
\begin{equation*}
M^{\Gamma}=\sqrt{\epsilon}\Bigl[M_{1}^{\Gamma}+{\epsilon}M_{2}^{\Gamma}+...\Bigr],\eqno{(A1)}
\end{equation*}
\begin{equation*}
M^{z}=M_{0}+{\epsilon}M_{1}^{z}+{\epsilon}^{2}M_{2}^{z}+...,\eqno{(A2)}
\end{equation*}
and
\begin{equation*}
B^{\Gamma}=\sqrt{\epsilon}\Bigl[B_{1}^{\Gamma}+{\epsilon}B_{2}^{\Gamma}+...\Bigr],\eqno{(A3)}
\end{equation*}
\begin{equation*}
B^{z}=B_{0}+{\epsilon}B_{1}^{z}+{\epsilon}^{2}B_{2}^{z}+...,\eqno{(A4)}
\end{equation*}
where $\Gamma=x,y$. To account for the conservation of length of the magnetization, consider the magnetization as
\begin{equation*}
{\bf{M}}=\vec{i}M^{x}+\vec{j}M^{y}+\vec{k}M^{z},\eqno{(A5)}
\end{equation*}
\begin{equation*}
{\bf{M^2}}=(M^{x})^{2}+(M^{y})^{2}+(M^{z})^{2},\eqno{(A6)}
\end{equation*}
where $\vec{i}$, $\vec{j}$ and $\vec{k}$ are unit vectors. After substituting the non-uniform expansions Eqs. (A1 $\&$ A2) in Eq. (A6), we get
\begin{equation*}
{\bf{M^2}}={M_{0}}^{2}+{\epsilon}\Bigl[2M_{0}M_{1}^{z}+(M_{1}^{x})^{2}+(M_{1}^{y})^{2}\Bigr]+...\eqno{(A7)}
\end{equation*}
Collecting the coefficients of $\epsilon$ at different orders of $\epsilon$ since $\bf{M}^{2}=1$,\\
$\epsilon^{0}$ ~~~:~ ${M_{0}}^{2}=1$ \\
and\\
$\epsilon^{1}$ ~~~:~ $2M_{0}M_{1}^{z}+(M_{1}^{x})^{2}+(M_{1}^{y})^{2}=0$.\\
On solving the above equation, we obtain
\begin{equation*}
M_{1}^{z}=-{1\over{2M_{0}}}\Bigl[(M_{1}^{x})^{2}+(M_{1}^{y})^{2}\Bigr].\eqno{(A8)}
\end{equation*}
By making use of the expression $\psi=M_{1}^{x}-i M_{1}^{y}$, we get
\begin{equation*}
M_{1}^{z}=-{1\over{2M_{0}}}\Bigl[(M_{1}^{x})^{2}+(M_{1}^{y})^{2}\Bigr]=-{1\over{2M_{0}}}|\psi|^{2}. \eqno{(A9)}
\end{equation*}
Similarly, using the complex field $\psi$, the components $M_{1}^{x}$ and $M_{1}^{y}$ of the magnetization of the anisotropic ferromagnetic medium can be constructed as $M_{1}^{x}= \Bigl[{\psi+\psi^{*}\over{2}}\Bigr]$ and $M_{1}^{y}= -\Bigl[{\psi-\psi^{*}\over{2i}}\Bigr]$.

\end{document}